\documentstyle[12pt,epsfig,a4]{article} 
\newcommand{\vs}{\vspace{-0.25cm}}
\begin{document}
\begin{center}
{\Large{\bf Radiative corrections to real and virtual muon Compton 
scattering revisited}}
\bigskip

N. Kaiser\\
\medskip
{\small Physik-Department T39, Technische Universit\"{a}t M\"{u}nchen,
    D-85747 Garching, Germany}
\end{center}
\medskip
\begin{abstract}
We calculate in closed analytical form the one-photon loop radiative 
corrections to muon Compton scattering $\mu^- \gamma \to \mu^- \gamma $. 
Ultraviolet and infrared divergencies are both treated in dimensional 
regularization. Infrared finiteness of the (virtual) radiative corrections is 
achieved (in the standard way) by including soft photon radiation below an 
energy cut-off $\lambda$. We find that the anomalous magnetic moment 
$\alpha/2\pi$ provides only a very small portion of the full radiative 
corrections. Furthermore, we extend our calculation of radiative corrections 
to the muon-nucleus bremsstrahlung process (or virtual muon Compton scattering
$\mu^-\gamma_0^* \to \mu^- \gamma $). These results are particularly relevant 
for analyzing the COMPASS experiment at CERN in which muon-nucleus 
bremsstrahlung serves to calibrate the Primakoff scattering of high-energy 
pions off a heavy nucleus with the aim of measuring the pion electric and 
magnetic polarizabilities. We find agreement with an earlier calculation of 
these radiative corrections based on a different method.
\end{abstract}

\bigskip

PACS:  12.20.-m, 12.20.Ds, 14.70.Bh
\section{Introduction and summary}
At present there is much interest in a precise experimental determination of
the (charged) pion electric and magnetic polarizabilities, $\alpha_\pi$ and 
$\beta_\pi$. Within the systematic framework of chiral perturbation theory the
firm prediction, $\alpha_\pi-\beta_\pi= (5.7\pm1.0)\cdot 10^{-4}\,$fm$^3$ 
\cite{gasser}, has been obtained for the (dominant) polarizability difference 
at two-loop order. However, this chiral prediction is in conflict with the 
existing experimental determinations of $\alpha_\pi- \beta_\pi=(15.6\pm 7.8) 
\cdot 10^{-4}\,$fm$^3$ from Serpukhov \cite{serpukhov} and $\alpha_\pi-\beta_\pi 
=(11.6\pm 3.4)\cdot 10^{-4}\,$fm$^3$ from Mainz \cite{mainz}, which amount to 
values more than twice as large. Certainly, these existing experimental 
determinations of $\alpha_\pi-\beta_\pi$ raise doubts about their correctness 
since they violate the chiral low-energy theorem notably by a factor 2. 

In that contradictory situation it is promising that the ongoing COMPASS 
\cite{compass} experiment at CERN aims at measuring the pion polarizabilities 
with high statistics using the Primakoff effect. The scattering of high-energy 
negative pions in the Coulomb field of a heavy nucleus (of charge $Z$) gives 
access to cross sections for $\pi^- \gamma$ reactions through the equivalent
photons method \cite{pomer}. In practice, one analyzes the 
spectrum of bremsstrahlung photons produced in the reaction $\pi^- Z\to\pi^- 
Z\gamma$ in the so-called Coulomb peak. This kinematical regime is 
characterized by very small momentum transfers to the nuclear target such that 
virtual pion Compton scattering $\pi^- \gamma^*_0 \to\pi^- \gamma$ occurs as the 
dominant subprocess (in the one-photon exchange approximation). The deviations 
of the measured spectra (at low $\pi^-\gamma$ center-of-mass energies) from 
those of a point-like pion are then attributed to the pion structure as 
represented by its electric and magnetic polarizabilities \cite{serpukhov} 
(taking often $\alpha_\pi+\beta_\pi=0$ as a constraint). It should
be stressed here that the systematic treatment of virtual pion Compton 
scattering $\pi^- \gamma^*_0 \to \pi^- \gamma$ in chiral perturbation theory 
yields at the same order as the polarizability difference $\alpha_\pi-\beta_\pi$ 
a further pion-structure effect in form of a unique pion-loop correction 
\cite{unkmeir,radcor} (interpretable as photon scattering off the 
``pion-cloud around the pion''). In the case of real pion Compton scattering 
$\pi^- \gamma \to \pi^- \gamma$ this loop correction compensates partly the 
effects from the pion polarizability difference \cite{picross}. A minimal 
requirement for improving future analyses of pion-nucleus bremsstrahlung 
$\pi^- Z\to\pi^- Z\gamma$ is therefore to include the loop correction 
predicted by chiral perturbation theory. At the required level of accuracy it 
is also necessary to include higher order electromagnetic corrections to the 
pion-nucleus bremsstrahlung process $\pi^-Z\to\pi^- Z\gamma$. These radiative 
corrections of order $\alpha$ have been calculated recently in 
ref.\cite{bremsrad}, where explicit analytical expressions for the 
corresponding one-photon loop amplitudes have been written down. Their 
implementation into the analysis of the COMPASS data is currently planned.

The setup of the COMPASS experiment \cite{compass} gives the possibility to 
switch (within a short time) from a pion beam to a muon beam which allows 
for precise acceptance studies with a point-like (purely electromagnetically 
interacting) particle. In this way the muon-nucleus bremsstrahlung process  
$\mu^- Z\to\mu^- Z\gamma$  serves the purpose to calibrate the Primakoff 
scattering events of high-energy pions off a heavy nucleus (of charge $Z$). 
Clearly, in order to preserve the level of accuracy one should take into 
account also the radiative corrections to the Bethe-Heitler cross section, 
which describes the muon-nucleus bremsstrahlung process $\mu^- Z\to\mu^- 
Z\gamma$ only at lowest order. The purpose of the present work is therefore 
to (re-)calculate the one-photon loop radiative corrections to muon-nucleus 
bremsstrahlung (or virtual muon Compton scattering). Radiative corrections to
bremsstrahlung have been computed already 50 years ago by Fomin in 
refs.\cite{fomin1,fomin2}. The analytical expression given in eq.(36) of 
ref.\cite{fomin2} (in terms of auxiliary functions defined by multiple 
integrals and derivatives thereof) is somewhat cumbersome for a numerical 
evaluation. Also, the limiting cases discussed explicitly in refs.\cite{fomin1,
fomin2} do not apply to the kinematics of the COMPASS experiment,
in which the muon beam and photon energy are very large, whereas the momentum 
transfer to the nucleus is very small. In this situation it is helpful to have 
an independent derivation of these radiative corrections based on a different 
calculational method as employed in ref.\cite{fomin2}. We are using here the 
dimensional regularization of ultraviolet and infrared divergent loop diagrams. 
In particular, we specify as a function of the five independent kinematical 
variables, the pertinent expressions for the interference terms between 
one-photon loop and tree diagrams summed over the muon spin and photon 
polarization states. In this detailed and explicit form our analytical results 
for the radiative corrections to virtual muon Compton scattering can be 
readily utilized for analyzing the muon Primakoff data of the COMPASS
experiment. 

The present paper is organized as follows. Section 2 is devoted to the (simpler)
calculation of the one-photon loop radiative corrections to real muon Compton 
scattering $\mu^- \gamma\to\mu^- \gamma$. The analogous case of electron 
Compton scattering has been considered long ago by Brown and Feynman in  
ref.\cite{feyn} and the final result for the radiative corrections has been
documented in the textbook on quantum electrodynamics by Akhiezer
and Berestetskii \cite{akhieser}. As a new element we use here dimensional 
regularization to treat both ultraviolet and infrared divergencies and we
write down in closed analytical form the expressions for the pertinent 
interference terms between one-photon loop and tree diagrams summed over the 
muon spin and photon polarization states. While the ultraviolet divergencies 
drop out in the renormalizable spinor electrodynamics at work, the infrared 
divergencies get removed (in the standard way) by including soft photon 
radiation below an energy cut-off $\lambda$. We confirm the absolute 
correctness of the final formula for these radiative corrections written in 
eq.(52.12) of the book by Akhiezer and Berestetskii \cite{akhieser}. In the 
closer analysis, we find that the muon anomalous magnetic moment $\kappa = 
\alpha/2\pi$ provides only a very small part of the (virtual) radiative 
corrections. Moreover, we compare these radiative corrections with those to 
(spin-0) pion Compton scattering and observe strong similarities concerning 
the energy and angular dependences. In section 3, we extend our calculation of 
radiative corrections to virtual muon Compton scattering $\mu^- \gamma^*_0\to
\mu^- \gamma$, where $\gamma^*_0$ denotes the virtual (Coulomb) photon that 
couples to the charge of a heavy nucleus in the bremsstrahlung process $\mu^-Z
\to \mu^- Z\gamma$. We derive again analytical expressions for the pertinent 
interference terms between one-photon loop and tree diagrams summed over the 
muon spin and photon polarization states. Since these expressions depend now 
on five independent kinematical variables, they become rather lengthy for some 
terms. The radiative correction factor arising from the soft photon radiation 
off the muon (which cancels the infrared divergencies due to the photon-loops) 
is evaluated together with all its finite pieces in the laboratory frame. For
this (soft photon) part the agreement with Fomin's earlier calculation 
\cite{fomin2} is obvious. For the (virtual) radiative corrections arising from 
photon loops the agreement between our and Fomin's calculation \cite{fomin2} 
can be verified numerically with good precision.   

The presentation of numerical results for the radiative corrections to 
the muon bremsstrah- lung process $\mu^- Z\to\mu^- Z \gamma$ in the COMPASS 
experiment is not pursued here, since that requires specification and 
implementation of the  actual experimental conditions (such as muon beam 
energy, constraints on detectable energy and angular ranges, etc.). The 
analytical results presented in this work form the basis for such a 
detailed study in the future.

\section{Radiative corrections to muon Compton scattering}
We start out with calculating the radiative corrections to (real) muon Compton
scattering. The in- and out-going four-momenta of the reaction, $\mu^-(p_1)+
\gamma(k_1,\epsilon_1)\to \mu^-(p_2)+ \gamma(k_2,\epsilon_2)$ give rise to the 
(Lorentz-invariant) Mandelstam variables:
\begin{eqnarray} && s=(p_1+k_1)^2=(p_2+k_2)^2= m^2(x+1)\,, \nonumber \\ && 
u=(p_1-k_2)^2 =(p_2-k_1)^2= m^2(y+1)\,,  \\ &&t=(p_1-p_2)^2 =(k_1-k_2)^2
=-m^2(x+y)\,, \nonumber  \end{eqnarray}
which obey the constraint $s+t+u = 2m^2$, with $m=105.658\,$MeV the muon mass. 
In the following it will be most advantageous to work (exclusively) with the 
two independent dimensionless variables $x$ and $y$ defined in eq.(1) in terms 
of $s$ and $u$. This pair of variables is preferred due to the presence of
poles in the tree level Compton diagrams (see Fig.\,1) which go simply as
$1/x$ and $1/y$ in these variables. In its general form, the T-matrix 
for Compton scattering off a spin-1/2 particle is parameterized by six 
independent (complex-valued) invariant amplitudes \cite{prange}, which are 
either even or odd under the crossing transformation $s \leftrightarrow u$. 
Observables, like the cross section, are then given by a quadratic form in 
these six amplitudes. We avoid the technical complication of decomposing the 
T-matrix of each individual diagram in the appropriate basis of 
Dirac-operators. Instead, we perform for all relevant contributions to the 
squared T-matrix directly the sums over the muon spin and photon polarization 
states (via a Dirac-trace and contractions with the metric tensor $g_{\mu\nu}$). 
In particular, for photon-loop diagrams the ($d$-dimensional) integration over 
the loop-momenta is carried out after the summation over the spin and 
polarization states. With that approach in mind, the unpolarized differential 
cross section for muon Compton scattering in the center-of-mass frame, 
including the radiative corrections of relative order $\alpha$, can be 
represented in following compact form: 
\begin{eqnarray} {d \sigma\over d\Omega_{\rm cm}}\!\!&=&\!\!{\alpha^2\over  2s} 
\bigg\{\bigg({A \over x}+{B \over y}\bigg) \otimes \bigg({A \over x}+{B \over 
y}\bigg) \nonumber \\ && +2\,{\rm Re}\bigg[\Big({\rm I+II+III+IV}\Big)\otimes 
\bigg({A \over x}+{B \over y}\bigg) +(x \leftrightarrow y)\bigg]\bigg\}\,.
\end{eqnarray}
with $\alpha = 1/137.036$ the fine-structure constant. Here, $A$ and $B$ 
denote the direct and crossed tree diagram (shown in Fig.\,1) and ${\rm I, II, 
III, IV}$ stand for the four classes of contributing one-photon 
loop diagrams (shown in Figs.\,2,3). The product symbol $\otimes$ designates 
the interference term between the T-matrices from two diagrams with the sums 
over the muon spin and photon polarization states already carried out. This 
multiple sum is efficiently done via a Dirac-trace and Lorentz-index 
contractions, ${1\over 8}{\rm tr}[(\slash\!\!\!p_2+m)O^{\mu\nu}(\slash\!\!\!p_1
+m) {\overline  O}_{\mu\nu}]$, where $O^{\mu\nu}$ denotes the Compton tensor. The 
symmetrization prescription $+(x \leftrightarrow y)$ in eq.(2) generates the 
additional contributions from the crossed one-photon loop diagrams (i.e. the 
diagrams in Figs.\,2,3 with crossed external photon lines).  
\begin{figure}
\begin{center}
\includegraphics[scale=1.,clip]{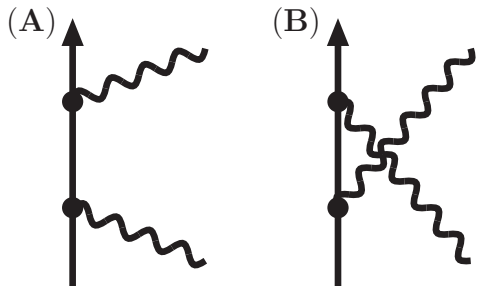}
\end{center}
\vspace{-.5cm}
\caption{Tree diagrams for muon Compton scattering.}
\end{figure}

The advantage of working with the dimensionless variables $x$ and $y$ shows up
already when evaluating the Born terms. The two tree diagrams shown in Fig.\,1 
lead to the following simple polynomial expressions: 
\begin{equation} A\otimes A = 2x-xy+4\,, \quad  A\otimes B=B\otimes A=x+y+4\,,
 \quad  B\otimes B =2y-xy+4\,. \end{equation}
When inserted into eq.(2), they produce the well-known Klein-Nishina cross 
section. In the center-of-mass frame the variables $x$ and $y$ are the related 
to the total center-of-mass energy $\sqrt{s}$ and the cosine of the scattering 
angle $\cos \theta_{\rm cm}$ in the following way:
\begin{equation} x =\bigg({\sqrt{s}\over m}\bigg)^2-1 \,, \qquad y = -{x \over
  2(1+x) }(2+x+x \cos \theta_{\rm cm}) \,. \end{equation} 
In the laboratory frame, on the other hand, $x$ and $y$ are directly 
proportional to the energies $E_\gamma$ and $E'_\gamma$ of the incoming and 
outgoing photon: 
\begin{equation} x = {2E_\gamma\over m}\,, \qquad  y = -{2E'_\gamma\over m}\,, 
\qquad  1-\cos\theta_{\rm lab}=-2\bigg({1\over x}+{1\over y}\bigg)\,,\end{equation}
with the scattering angle $\theta_{\rm lab}$ fixed by them. In the physical
region the inequalities, $x\geq 0$ and $ -x\leq y \leq -x/(1+x)\leq 0$, hold.
The radiative corrections to muon-pair annihilation $\mu^+ \mu^- \to \gamma
\gamma$ are obtained by specifying $x$ and $y$ in eq.(2) for that (crossed) 
reaction. 
\subsection{Evaluation of one-photon loop diagrams} 
In this section, we present analytical expressions (of order $\alpha$) for the 
interference terms between the one-photon loops diagrams and the tree diagrams 
for muon Compton scattering. We use the method of dimensional regularization 
to treat both ultraviolet and infrared divergencies (where the latter are
caused by the masslessness of the photon). The method consists in calculating
loop integrals in $d$ spacetime dimensions and expanding the results around
$d=4$. Divergent pieces of one-loop integrals generically show up in form of
the composite  constant:
\begin{equation} \xi = {1\over d-4} +{1\over2} (\gamma_E-\ln 4\pi )+\ln{m\over 
\mu}\,, \end{equation}
containing a simple pole at $d=4$. In addition, $\gamma_E = 0.5772\dots$ is the 
Euler-Mascheroni number and $\mu$ an arbitrary mass scale introduced in 
dimensional regularization in order to keep the mass dimension of the loop 
integrals independent of $d$. Ultraviolet (UV) and infrared (IR) divergencies 
are distinguished by the feature of whether the condition for convergence of 
the $d$-dimensional integral is $d<4$ or $d>4$. We discriminate them in the 
notation by putting appropriate subscripts, i.e. $\xi_{UV}$ and $\xi_{IR}$. In 
order to simplify all calculations, we employ the Feynman gauge where the 
photon propagator is directly proportional to the Minkowski metric tensor 
$g^{\mu\nu}$. Let us now enumerate the analytical results as they emerge from
the four classes of one-photon loop diagrams shown in Figs.\,2,3.

\begin{figure}
\begin{center}
\includegraphics[scale=1.,clip]{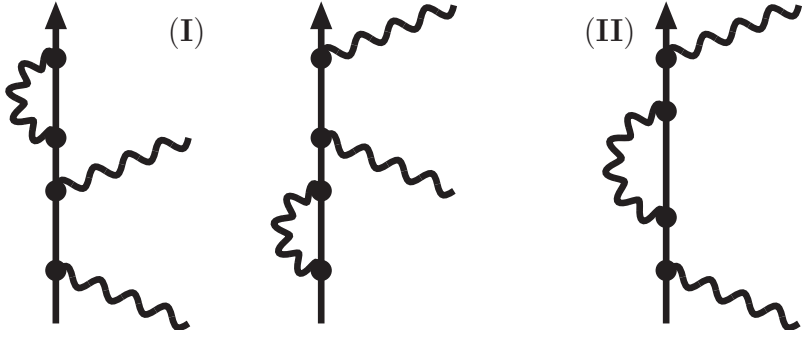}
\end{center}
\vspace{-.5cm}
\caption{One-photon loop diagrams, (I) and (II),  for muon Compton scattering.}
\end{figure}

\medskip
\noindent{\it Class I.} These photon-loop diagrams introduce the 
wavefunction renormalization factor $Z_2-1=\alpha(2\xi_{IR}+\xi_{UV}-2)/2\pi $ 
\cite{pokorski} of the muon and the pertinent interference terms with the 
(direct and crossed) tree diagrams read:  
\begin{equation} {\rm I} \otimes A ={\alpha \over 2\pi}\bigg\{(2 \xi_{IR}+
\xi_{UV}-2)\bigg( 2 -y +{4 \over x}\bigg)-y\bigg\}\,, \end{equation}
\begin{equation} {\rm I} \otimes B ={\alpha \over 2\pi}\bigg\{(2 \xi_{IR}+
\xi_{UV}-2)\, {x+y+4 \over x}-{y \over 2}\bigg\}\,. \end{equation}
The additive terms, $-y$ and $-y/2$, come from evaluating the Dirac-traces and  
Lorentz-index contractions consistently in $d$ spacetime dimensions\footnote{
The pole term in the ultraviolet divergence $\xi_{UV} = 1/(d-4) +\dots$ counts 
at this point.} and expanding the results around $d=4$. Within dimensional 
regularization it is crucial to follow this prescription in order to recover 
the correct low-energy limit $(k_1,k_2\to 0)$ for Compton scattering, namely 
the non-renormalization  of the Thomson amplitude \cite{radcor}.   

\medskip
\noindent{\it Class II.} This photon-loop diagram involves the off-shell 
fermion selfenergy subtracted by the mass shift. The corresponding
interference terms with the tree diagrams $A$ and $B$ read: 

\begin{eqnarray} {\rm II} \otimes A &=&{\alpha \over 4\pi x}\bigg\{2\xi_{UV}
(2x-xy+4)-xy-5x+{3x \over x+1}\nonumber \\ &&+\bigg[11x-x y+3y-26+{13-3y 
\over x+1}-{3\over (x+1)^2}\bigg] \ln(-x)\bigg\} \,,\end{eqnarray}

\begin{eqnarray} {\rm II} \otimes B &=&{\alpha \over 2\pi x}\bigg\{\xi_{UV}
(x+y+4) -{x y\over 2}+{x(y-1) \over x+1}  \nonumber \\ &&+\bigg[{3y-7\over x+1}
+{1-y \over (x+1)^2}-x-4y-2 \bigg] \ln(-x)\bigg\} \,.\end{eqnarray}

\begin{figure}
\begin{center}
\includegraphics[scale=1.,clip]{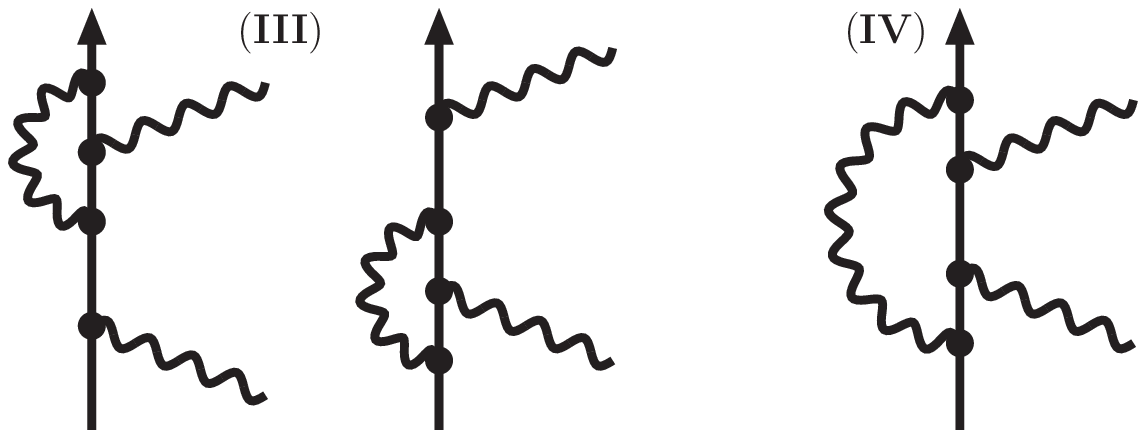}
\end{center}
\vspace{-.5cm}
\caption{One-photon loop diagrams, (III) and (IV),  for muon Compton 
scattering.}
\end{figure}
\medskip
\noindent{\it Class III.} These photon-loop diagrams involve the half 
off-shell vertex correction. Both of them contribute equally and the pertinent 
interference terms with the tree diagrams read: 

\begin{eqnarray} {\rm III} \otimes A &=&{\alpha \over 2\pi x}\bigg\{2\xi_{UV}
(x y-2x-4)+2x^2+4x y+7x+15+{1\over x+1}\nonumber \\ &&+ \bigg[7-10x-x^2-2y(x+1)
+{2y-8 \over x+1}+{1\over (x+1)^2}\bigg] \ln(-x)\nonumber \\ &&-4\bigg(2+x+y
+{4\over x}\bigg)\bigg[{\rm Li}_2(x+1)-{\pi^2\over 6}\bigg] \bigg\} \,,
\end{eqnarray}

\begin{eqnarray} {\rm III} \otimes B &=&{\alpha \over 2\pi x}\bigg\{-2\xi_{UV}
(x+y+4)-3y^2-2x y-8y-{2y\over x}(8+y)-{x^2\over x+1}\nonumber \\ &&+ \bigg[4x+
3x y+13y+3y^2+5-{1\over (x+1)^2}+{12-5y \over x+1}+{2y \over x}(8+y)\bigg] 
\ln(-x) \nonumber \\ && +\bigg[3y+1+{2y \over x}(3+y)+{y \over x^2}(8+y)\bigg]
\bigg[2{\rm Li}_2(x+1)-{\pi^2\over 3}\bigg] \bigg\} \,,\end{eqnarray}
where Li$_2(x) = \sum_{n=1}^\infty n^{-2} x^n = x \int_1^\infty d\xi [\xi(\xi-x)]^{-1}
\ln \xi$ denotes the conventional dilogarithmic function. A particular part 
of the vertex corrections (on the mass-shell) is the muon anomalous magnetic 
moment $\kappa = \alpha/2\pi$. Taken by its own  the anomalous magnetic 
moment leads (via an appropriate derivative-coupling vertex) to the 
contributions:

\begin{equation} \kappa \otimes A = -{3 \alpha \over 4\pi} (2x+y)\,, \qquad 
 \kappa \otimes B = {\alpha\, y\over 4\pi x}(4x+y)\,. \end{equation}
By comparing with the complete (off-shell) vertex corrections written in 
eqs.(11,12) one notices that there is no obvious way to localize  therein 
the anomalous magnetic moment contributions. On-shell and off-shell effects 
from the photon-loop appear to be inseparably mixed in the interference terms 
III$\,\otimes\, A$ and  III$\,\otimes\, B$.  

\medskip
\noindent{\it Class IV.} The evaluation the photon-loop box diagram is 
most challenging and one finds for its interference terms with the tree 
diagrams $A$ and $B$:
\begin{eqnarray} {\rm IV} \otimes A &=&{\alpha \over 2\pi }\Bigg\{
{(8+3x)y \over x+y}-4-4y+\bigg[x+2y+{3x\over x+1}\bigg]\ln(-x)\nonumber \\ &&
+\bigg[2+y+{4\over x}(y-1)\bigg]\bigg[{\rm Li}_2(x+1)-{\pi^2\over 6}\bigg] 
+4 \xi_{IR} \bigg(y-2-{4\over x}\bigg) \nonumber \\&&\times {2+x+y \over 
\sqrt{4+x+y}}\, L(x+y) +{2L(x+y)\over \sqrt{4+x+y}} \bigg[y-x-x^2-2xy+{4x(4-y)
\over x+y} \nonumber \\ &&+\bigg(x y+y^2-2x-12-{8\over x}(2+y)\bigg) \ln(-x)
\bigg]+\bigg(4-2y-x y-y^2-{4x y \over x+y}\bigg) \nonumber \\ &&\times  L^2(x+y)
+ {2+x+y\over h_+-h_-}\bigg\{\bigg(y-2-{4\over x}\bigg)\bigg[{\rm Li}_2(w)-
{\rm Li}_2(1-w) \nonumber \\ &&+{1\over 2}\ln^2w -{1\over 2}\ln^2(1-w)\bigg]+
\bigg({4\over x}+1-{y \over 2}\bigg) \Big[{\rm Li}_2(h_+)-{\rm Li}_2(h_-)
\Big]\bigg\} \Bigg\} \,,\end{eqnarray}

\begin{eqnarray} {\rm IV}\otimes B &=&{\alpha\over 2\pi}\Bigg\{{x\over x+1}
+{4y\over x}(y-1)+{y\over x+y}(8+x-2y) \nonumber \\ &&+\bigg[ {4y\over x}(1-y)
-4y-1-{1\over (x+1)^2}+{6-4y\over x+1}\bigg] \ln(-x)\nonumber \\ &&+\bigg[ 
x+y-1-{4+y\over x}+{2y\over x^2}(1-y)\bigg] \bigg[2{\rm Li}_2(x+1)-{\pi^2\over 
3}\bigg] \nonumber \\ && -{4 \xi_{IR} \over x}(2+x+y) \sqrt{4+x+y}\, L(x+y)+
{2L(x+y)\over \sqrt{4+x+y}}\nonumber \\ &&\times  \bigg[x+7y-16+2x y+3y^2+{4x
(4-y)\over x+y}+\bigg(2(x+y)^2-24 \nonumber \\ &&-{2\over x}(8+6y+y^2)\bigg) 
\ln(-x)\bigg] +\bigg(4+2x-2(x+y)^2+{4y^2\over x+y}\bigg)\nonumber \\ && \times
L^2(x+y) + {1\over x(h_+-h_-)}\bigg\{(2+x+y)(4+x+y)\nonumber \\ &&\times 
\bigg[{\rm Li}_2(1-w)-{\rm Li}_2(w) \nonumber +{1\over 2}\ln^2(1-w) -{1\over 2}
\ln^2w \bigg]\nonumber \\ &&+ \Big(8+12x-x^3+2y(3-x^2)+y^2(1-x)\Big) \Big[
{\rm Li}_2(h_+)-{\rm Li}_2(h_-)\Big]\bigg\} \Bigg\} \,.\end{eqnarray}
Here, we have introduced the frequently occurring logarithmic loop function: 
\begin{equation} L(x)= {1\over \sqrt{x}} \ln{ \sqrt{4+x}+\sqrt{x}\over 2}\,, 
\end{equation}
and the auxiliary variables: 
\begin{equation} w ={1\over 2} \bigg( 1-\sqrt{{x+y\over 4+x+y}}\,\bigg)\,, 
\qquad h_\pm = {1\over  2}\Big[-x-y\pm\sqrt{(x+y)(4+x+y)}\,\Big]\,.\end{equation}
It is remarkable that only one single loop integral involving four propagators  
contributes at the end to the interference terms IV$\,\otimes\, A$ and IV$\,
\otimes\, B$. This (scalar) loop integral is infrared divergent and it can 
actually be solved in terms of logarithms ($ L(x+y),\ln(-x), \ln^2w,
\ln^2(1-w)$) and dilogarithms (${\rm Li}_2(1-w),{\rm Li}_2(w), 
{\rm  Li}_2(h_\pm)$). As an immediate check, one verifies that the ultraviolet 
divergent terms proportional to $\xi_{UV}$ cancel out in the sums 
(I+II+III)$\,\otimes\, A$ and (I+II+III)$\,\otimes\, B$.

\subsection{Infrared finiteness}
In the next step we have to consider the infrared divergent terms proportional 
to $\xi_{IR}$. Inspection of eqs.(7,8,14,15) reveals that these scale with the 
Born terms $A \otimes A/x$ and $A \otimes B/x$  given in eq.(3). As 
a consequence of that feature, the infrared divergent loop corrections 
multiply differential cross section $d\sigma/d\Omega_{\rm cm}$ at 
leading order by a ($x \leftrightarrow y$ crossing-symmetric) factor:  
\begin{equation} \delta_{\rm virt}^{(\rm IR)}= {2 \alpha \over \pi} \Bigg[ 1-
{4+2x+2y\over \sqrt{4+x+y}}\, L(x+y) \Bigg] \, \xi_{IR} \,. \end{equation}
The (unphysical) infrared divergence $\xi_{IR}$ gets canceled at the level of 
the (measurable) cross section by the contributions of soft photon 
bremsstrahlung. In its final effect, the (single) soft photon radiation
multiplies the tree level cross section $d\sigma/d\Omega_{\rm cm}$ by a 
factor \cite{radcor,radiat}: 
\begin{equation} \delta_{\rm soft}= \alpha\, \mu^{4-d}\!\!\int\limits_{|\vec 
l\,|<\lambda} \!\!{d^{d-1}l  \over (2\pi)^{d-2}\, l_0} \bigg\{ {2p_1\cdot p_2 
\over p_1 \cdot l \, p_2 \cdot l} - {m^2 \over (p_1 \cdot l)^2} - {m^2 \over 
(p_2 \cdot l)^2} \bigg\} \,, \end{equation}
which depends on a small energy cut-off $\lambda$. Working out this
momentum space integral by the method of dimensional regularization (with 
$d>4$) one finds that the infrared divergent correction factor $\delta_{\rm  virt}
^{(\rm IR)} \sim \xi_{IR}$ in eq.(18) gets eliminated and the following finite 
radiative  correction factor remains:
\begin{eqnarray}\delta_{\rm real}^{(\rm cm)}&=&{\alpha \over \pi}\Bigg\{\bigg[ 2- 
{8+4x+4y \over\sqrt{4+x+y}}\, L(x+y) \bigg] \ln{m \over 2\lambda}+{x+2\over x}
\ln(x+1)\nonumber \\ && \qquad - \int_0^{1/2}\!\! d\tau \,{(x+2)(2+x+y)\over 
[1+(x+y) \tau(1-\tau)]\sqrt{W} } \ln{x+2+\sqrt{W} \over  x+2-\sqrt{W} }\Bigg\}
\,,    \end{eqnarray}
with the abbreviation $W = x^2 - 4(x+1)(x+y)\tau(1-\tau)$. We note that
the terms beyond those proportional to $\ln(m/2\lambda)$ are specific for 
the evaluation of the soft photon correction factor $\delta_{\rm soft}$ in the 
$\mu^-\gamma$ center-of-mass frame with $\lambda$ an infrared cut-off therein.

Having the analytical results at hand, one can show that both the virtual and 
the real radiative corrections vanish as one approaches (at fixed scattering 
angle $\theta_{\rm cm}$) the Compton threshold, $\sqrt{s}\to m$ or $x\to 0$. 
This non-renormalization of the Thomson cross section $d\sigma /d\Omega_{\rm cm}
= \alpha^2(1+\cos^2\theta_{\rm cm})/2m^2$ serves as an important check
 on the method of dimensional regularization where spin and polarization sums 
are extended from $4$ to $d$ spacetime dimensions. 

\subsection{Results and discussion}
We are now in the position to present numerical results for the radiative
corrections to muon Compton scattering $\mu^- \gamma \to \mu^- \gamma$. The
complete radiative correction factor is $\delta_{\rm real}$ written in eq.(20)
plus the sum of all interference terms (second line in eq.(2)) divided by the 
Born terms (first line in eq.(2)). First, it is important to note that our
calculation (which employs dimensional regularization to treat both ultraviolet 
and infrared divergent loop integrals) confirms the absolute correctness of 
the expression written in eq.(52.12) of ref.\cite{akhieser} for the radiative
corrections to spin-1/2 Compton scattering.  Fig.\,4 shows in percent the 
total radiative correction factor for four selected center-of-mass energies 
$\sqrt{s} = (2,3,4,5)m$. The detection threshold for soft photons has been set 
to the value $\lambda = 3.8\,$MeV.\footnote{For a meaningful comparison of the 
radiative corrections to muon and pion Compton scattering the ratio of 
infrared cut-off $\lambda$ to the particle mass has to be chosen equally.} One 
observes that the radiative corrections become maximal in backward directions 
$\cos \theta_{\rm cm} \simeq -1$, reaching values up to $-3\%$ at $\sqrt{s}= 5m$. 
In Fig.\,5 we show for comparison the radiative corrections to pion Compton
scattering \cite{radcor} (for an equivalent infrared cut-off of $\lambda=
5\,$MeV). One notices that the overall magnitude and angular dependence of the 
radiative corrections are very similar in both cases. Interestingly, the dip in 
backward directions is more pronounced in the spin-0 case.     
\begin{figure}
\begin{center}
\includegraphics[scale=0.45,clip]{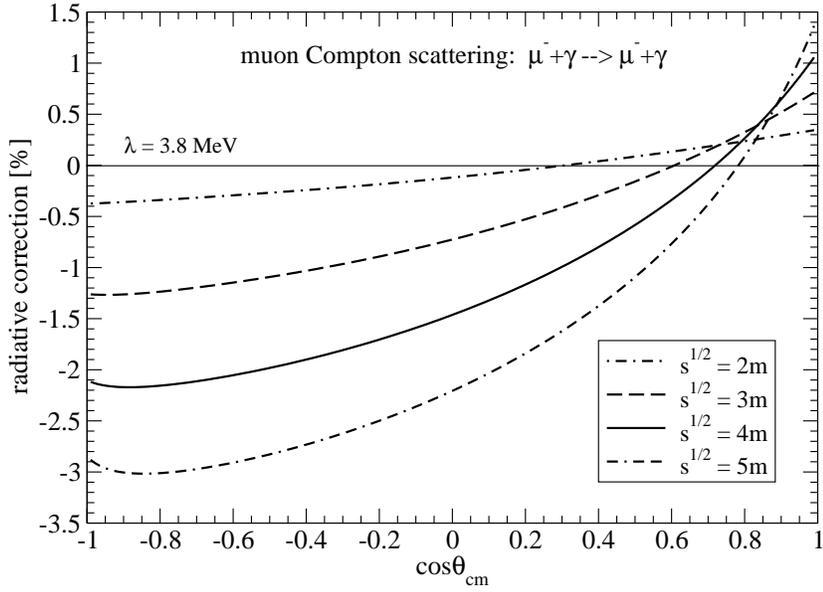}
\end{center}
\vspace{-0.9cm}
\caption{One-photon loop radiative corrections to muon Compton scattering.}
\end{figure}

\begin{figure}
\begin{center}
\includegraphics[scale=.45,clip]{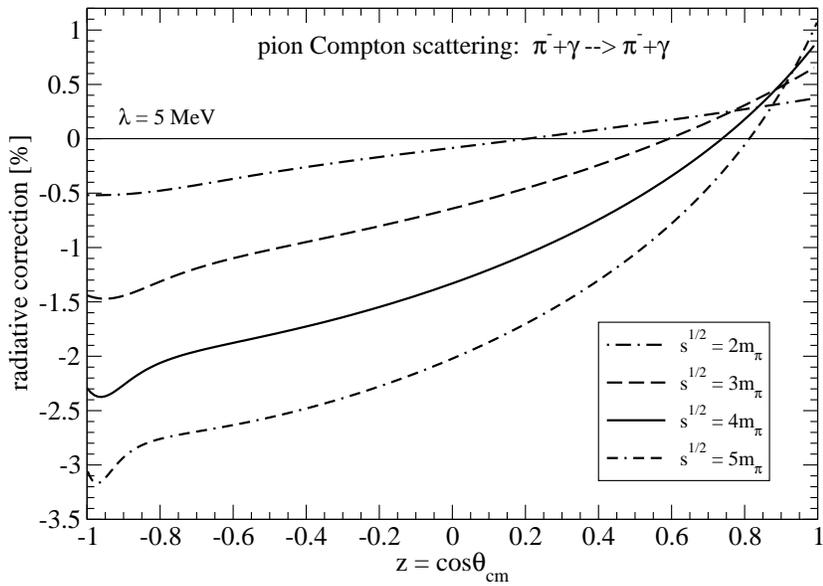}
\end{center}
\vspace{-0.9cm}
\caption{Radiative corrections for pion Compton scattering $\pi^- \gamma \to
\pi^- \gamma$.}
\end{figure}

\begin{figure}
\begin{center}
\includegraphics[scale=0.45,clip]{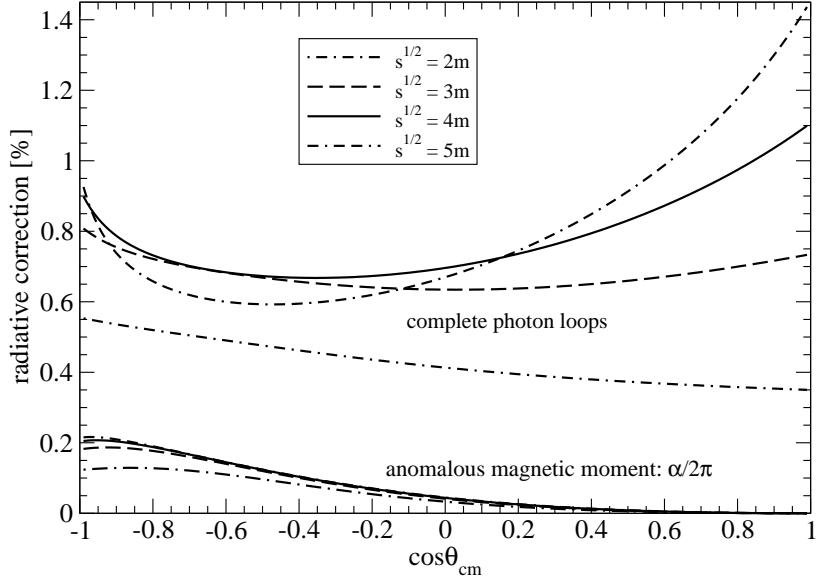}
\end{center}
\vspace{-0.9cm}
\caption{Effect of anomalous magnetic moment $\kappa = \alpha/2\pi$
 compared to the complete photon-loops.}
\end{figure}

Next, we discuss the role of the muon anomalous magnetic moment $\kappa=
\alpha/2\pi$. As a radiative correction it represents the on-shell part of the 
vertex corrections generated by the photon-loop diagrams of class III (see 
Fig.\,3). The comparison made in Fig.\,6 informs us that the muon anomalous 
magnetic moment $\kappa= \alpha/2\pi$ provides only a very small portion of 
the complete radiative corrections arising from the full set of photon-loop 
diagrams. Most noticeably, the contribution of the anomalous magnetic moment 
vanishes in forward directions $\cos\theta_{\rm cm} \simeq 1$, whereas the
complete radiative corrections grow strongly with the center-of-mass energy 
$\sqrt{s}$ in that region. Note that we have omitted the soft photon
contribution $\delta_{\rm real}^{(\rm cm)}$ in the comparison shown in Fig.\,6.

The radiative corrections in forward direction $\cos\theta_{\rm cm} =1$ are of
particular interest, because for this kinematical situation the soft
bremsstrahlung contribution $\delta_{\rm real}^{(\rm cm)}$ vanishes identically. The 
soft photon emission before and after the Compton scattering process interfere 
in this case destructively with each other. After taking carefully the forward 
limit $y\to -x$ of all terms calculated in section 2.1, we get for the 
radiative corrections to muon Compton scattering in forward direction:
\begin{eqnarray} \delta_0(x) &=&{\alpha\over 2\pi} \bigg\{{4-3x^2\over x^2-1}+
{x^4-3x^2 \over (x^2-1)^2} \ln x+{\pi^2\over 3x^2}(x^2-8) \nonumber \\ && 
+ {8+4x-x^2 \over x^2}\, {\rm Re} \,{\rm Li}_2(x+1) + {8-4x-x^2\over x^2}\,
{\rm Li}_2(1-x) \bigg\} \,.\end{eqnarray}
The dependence of this function on the center-of-mass energy $\sqrt{s}= m 
\sqrt{x+1}$ is reproduced by the full curve in Fig.\,7. Its asymptotic 
behavior is $(\alpha/2\pi) \ln^2x$. The dashed curve in Fig.\,7 shows for 
comparison the same quantity for Compton scattering off a (point-like) spin-0 
particle. Using the analytical results in section 3 of ref.\cite{radcor} one
deduces the expression: 
\begin{equation} \delta_0(x)^{\rm spin-0} ={2\alpha\over \pi} \bigg\{{x^2\ln x 
\over x^2-1} -{2\pi^2\over 3x^2}-1 + {2+x\over x^2}\, {\rm Re} \,{\rm Li}_2(x+1) 
+ {2-x\over x^2}\,{\rm Li}_2(1-x) \bigg\} \,,\end{equation}
for the radiative corrections to pion Compton scattering in forward direction.  
The asymptotic behavior $(2\alpha/\pi) \ln x$ in the spin-0 case turns out to
be weaker than for the muon. Presumably, this difference has its origin in
the additional magnetic coupling of the photon to a (point-like) spin-1/2 
particle.    
\begin{figure}
\begin{center}
\includegraphics[scale=0.45,clip]{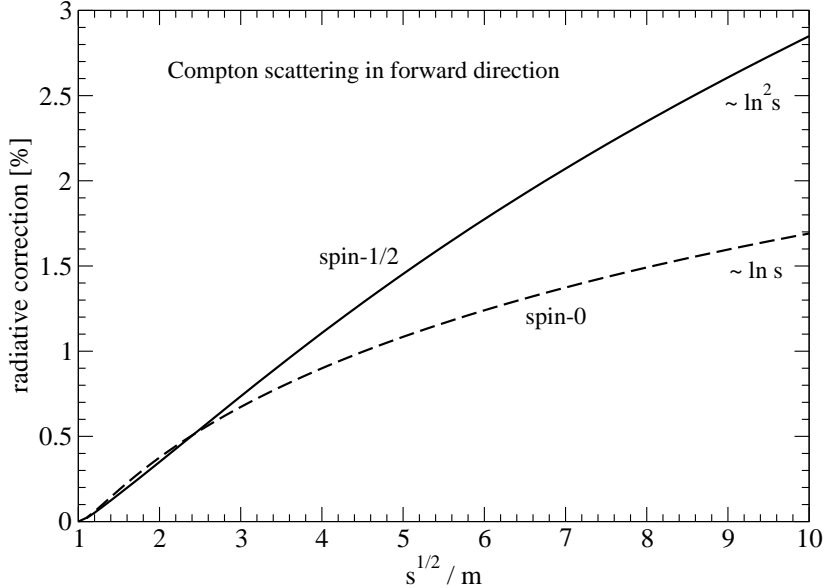}
\end{center}
\vspace{-0.9cm}
\caption{Radiative correction to muon and pion Compton scattering in forward 
direction.}
\end{figure}

\section{Radiative corrections to muon-nucleus bremsstrahlung}
In this section we extend our calculation of radiative corrections to the 
muon-nucleus brems- strahlung process $\mu^- Z \to \mu^- Z \gamma$. In the 
one-photon exchange approximation a virtual Coulomb photon couples to 
the heavy nucleus of charge $Z$ and the bremsstrahlung process is governed
by the virtual muon Compton scattering: $\mu^-(p_1)+\gamma^*_0(q) \to \mu^-(p_2)
+\gamma(k,\epsilon)$. The in- and out-going four-momenta of that subprocess 
give rise to the following kinematical variables:
\begin{equation} s=(p_1+q)^2=(p_2+k)^2= m^2(x+1)\,, \qquad u=(p_1-k)^2  =(p_2-q)^2
= m^2(y+1)\,, \end{equation}
\begin{equation}E_1=m\,e_1\,,\qquad E_2=m\,e_2\,,\qquad q^2=-m^2 Q\,,\qquad
  q_0=0\,, \end{equation}
where we have also noted that the extremely small recoil energy $-q_0 = 
\vec q\,^2/2M_{\rm nucl} \approx 0 $ of the nucleus can be neglected. In the 
following it will be most advantageous to work with the five independent 
dimensionless variables $e_1,e_2, x, y, Q$. The unpolarized (fivefold) 
differential cross section for muon-nucleus bremsstrahlung in the laboratory 
frame reads \cite{itzykson}: 
\begin{equation} {d^5 \sigma\over d\omega d\Omega_{\gamma} d\Omega_{\mu}} = {Z^2
\alpha^3 \omega |\vec p_2|\over\pi^2|\vec p_1||\vec q\,|^4}\,H\,,\end{equation}
with $\omega = k_0 =m(e_1-e_2)$, $|\vec p_j| = m \sqrt{e_j^2-1}$, $|\vec q\,|= m
\sqrt{Q}$ and $H$ is the squared amplitude summed over muon spin and photon 
polarization states. This sum is carried out most conveniently via a 
Dirac-trace and a contraction with the (negative) Minkowski metric tensor, 
$-{1\over 8}{\rm tr}[(\slash\!\!\!p_2+m)O^{\nu} (\slash\!\!\!p_1+m) {\overline 
O}_{\nu}]$. In that procedure the virtual Coulomb photon is given the
polarization four-vector $v^\mu =(1,\vec 0\,)$ which provides in the form of 
Lorentz-scalar products the muon energies $E_1= m\,e_1$ and $E_2= m\,e_2$ as
further independent variables.      

\begin{figure}
\begin{center}
\includegraphics[scale=1.,clip]{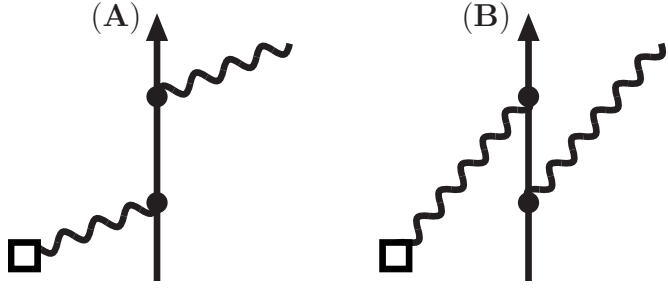}
\end{center}
\vspace{-.5cm}
\caption{Tree diagrams for virtual muon Compton scattering.}
\end{figure}
The direct and crossed tree diagrams for virtual muon Compton scattering 
(named again $A$ and $B$) are shown in Fig.\,8. The open square in the left
lower corner should symbolize the heavy nucleus to which the virtual 
Coulomb photon $\gamma_0^*$ couples. The contribution of these two tree 
diagrams to the squared amplitude $H$ has the form:
\begin{equation}H_{\rm tree} =\bigg({A \over x}+{B \over y}\bigg) \otimes 
\bigg({A \over x}+{B \over  y}\bigg)\,,   \end{equation}
with 
\begin{eqnarray}&& A\otimes A=2e_1^2(x-2)-2e_1e_2\,x+Q+{x y\over 2}  \,,\\ && 
A\otimes B=e_1e_2(x+y-4)-e_1^2(Q+y)-e_2^2(Q+x)+{Q\over 2}(2+Q+x+y) \,, \\ &&  
B\otimes B=2e_2^2(y-2)-2e_1e_2\,y+Q+{x y\over 2}\,,\end{eqnarray}
where the product symbol $\otimes$ designates again the interference term 
between (the T-matrices from) two diagrams with the sums over muon spins and 
photon polarizations already carried out. As it is written here in eqs.(26-29) 
the term $H_{\rm  tree}$ reproduces to the familiar Bethe-Heitler cross section 
\cite{itzykson} (now reexpressed in terms of the independent variables $e_1,e_2,
x, y, Q$). We note as an aside that for a spin-0 particle (e.g. a pion) 
$H_{\rm  tree}$ reduces to a much simpler form:
\begin{equation} H_{\rm  tree}^{(\rm spin-0)} = -4 \bigg( {e_1\over x}+
  {e_2\over y} \bigg)^2 - {4e_1 e_2 Q\over x y} -1\,.  \end{equation}

The radiative corrections of order $\alpha$ to virtual muon Compton scattering
$\mu^- \gamma^*_0 \to \mu^- \gamma $ can be summarized by the contribution: 
 \begin{equation}H_{\rm loop} =2\, {\rm Re}\bigg[\Big({\rm I+II+III+IV+V}\Big)
\otimes \bigg({A \over x}+{B \over  y}\bigg)+(x \leftrightarrow y, \, e_1 
\leftrightarrow-e_2 )\bigg]\,,  \end{equation} 
where I, II, III, IV, V  denote the five classes of one-photon loop diagrams
shown in Figs.\,9,10. The symmetrization prescription $+(x \leftrightarrow y, 
\, e_1 \leftrightarrow-e_2 )$ supplies the additional contributions from the 
crossed one-photon loop diagrams. These are obtained by interchanging the 
coupling vertex of the left and the right photon line in the diagrams of 
Figs.\,9,10.

\subsection{Evaluation of one-photon loop diagrams} 
In this section we collect the analytical expressions for the interference
terms between one-photon loop diagrams and tree diagrams for virtual muon 
Compton scattering. Such a direct (diagrammatic) calculation of the radiative 
corrections to bremsstrahlung has been deferred in ref.\cite{fomin1,fomin2} as 
being too difficult and instead the so-called mass operator method
\cite{newton} has been used.

\begin{figure}
\begin{center}
\includegraphics[scale=1.,clip]{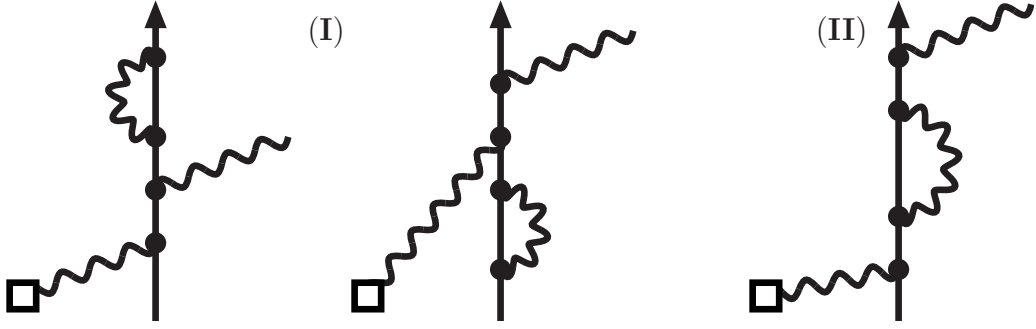}
\end{center}
\vspace{-.5cm}
\caption{One-photon loop diagrams, (I) and (II), for virtual muon Compton 
scattering.}
\end{figure}

\medskip
\noindent{\it Class I.} These photon-loop diagrams introduce the wavefunction 
renormalization factor $Z_2-1$ and the pertinent interference terms 
with the tree diagrams ($A$ and $B$) read:
\begin{equation} {\rm I}\otimes A ={\alpha \over 2\pi}\bigg\{(2\xi_{IR}+\xi_{UV}
-2){ A\otimes A\over x}+e_1^2-e_1 e_2+{y\over 4}\bigg\}\,, \end{equation}
\begin{equation} {\rm I}\otimes B ={\alpha \over 4\pi x}\bigg\{2(2\xi_{IR}+
\xi_{UV}-2) A\otimes B+ e_1e_2(x+y)-e_1^2\,y-e_2^2\,x -Q(e_1-e_2)^2+{xy\over  2}
\bigg\} \,, \end{equation}
with  $A\otimes A$ and  $A\otimes B$ given in eqs.(27,28).

\medskip
\noindent{\it Class II.} This photon-loop diagram involves the off-shell 
selfenergy subtracted by the mass shift and the corresponding
interference terms with the tree diagrams read: 
\begin{eqnarray}{\rm II}\otimes A &=& {\alpha \over 4\pi x}\bigg\{2\xi_{UV}
A \otimes A+{x\over x+1}\bigg(2e_1^2-{Q\over 2}+x\bigg)+\Big[4e_1^2(8+8x+x^3) 
\nonumber \\ && +(4e_1 e_2-y)(2x+x^2-x^3)-Q(8+10x+x^2)-4x^2-6x^3\Big]
{\ln(-x) \over 2(x+1)^2} \bigg\}\,,\end{eqnarray}
\begin{eqnarray}{\rm II}\otimes B &=& {\alpha \over 4\pi x}\bigg\{2\xi_{UV}
A \otimes B +{1\over 2(x+1)}\Big[2e_1e_2(2Q+x+2Qx+y)-2e_1^2(Q+y) \nonumber \\ &&
+2e_2^2(x-Q)-Q x(3+Q+x+y)+ x^2y\Big] +\bigg[ {x y\over2}(3x+2) \nonumber \\ && 
-e_2^2x^2(x+2) +\bigg( {2e_1^2-Q\over 2}(Q+y)-e_1e_2 y+e_2^2Q\bigg)(4+4x-x^2) 
\nonumber \\ && +e_1 e_2(16+20x+4x^2+x^3) +{Q\over 2} (x^3+x^2-10x-8)\bigg]
{\ln(-x) \over (x+1)^2}\bigg\}\,.\end{eqnarray}

\begin{figure}
\begin{center}
\includegraphics[scale=1.,clip]{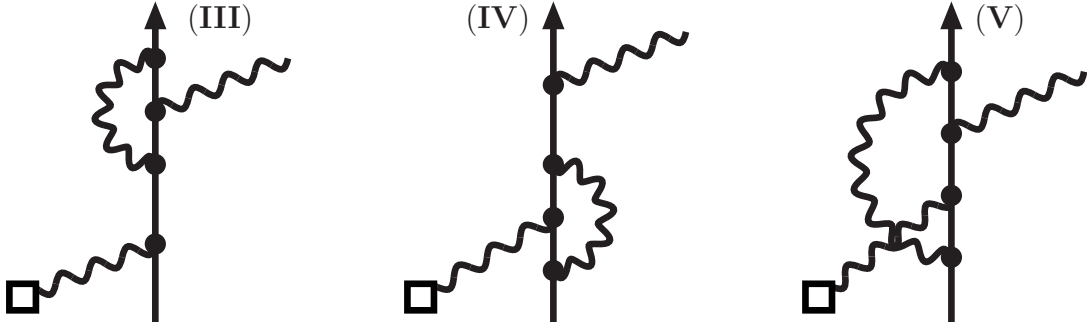}
\end{center}
\vspace{-.5cm}
\caption{One-photon loop diagrams, (III), (IV) and (V), for virtual muon 
Compton  scattering.}
\end{figure}

%\medskip
\noindent{\it Class III.} The vertex correction to the (upper) real photon
emission introduces loop functions that depend only on the variable $x$
(related to $s = (p_2+k)^2$). The pertinent interference terms with the tree 
diagrams ($A$ and $B$) read:
\begin{eqnarray}{\rm III}\otimes A &=& {\alpha \over 4\pi x}\bigg\{-2\xi_{UV}
A \otimes A -x^2 -18e_1^2 + {9Q+1 \over 2} +{4e_1^2-1-Q \over 2(x+1)}\nonumber 
\\ &&+{x\over 2}(12e_1 e_2-4e_1^2-2Q -3y-1)+ \bigg[Q(x^2-2x-2)+x^3+6x^2 \nonumber
 \\ && +4x +4e_1^2(x^2+8x+6)+(2y-8e_1e_2)(x^2+3x+2)\bigg] {x \ln(-x) \over 
2(x+1)^2} \nonumber \\ &&+ \bigg[ 8e_1^2-8e_1 e_2+x+2y +{4\over x}(4e_1^2-Q) 
\bigg]\bigg[{\rm Li}_2(x+1)-{\pi^2\over 6} \bigg]  \bigg\}\,,\end{eqnarray}

\begin{eqnarray}{\rm III}\otimes B &=& {\alpha \over 4\pi x}\bigg\{-2\xi_{UV}
A \otimes B + {x\over 2}(1+Q+2y-2e_1 e_2)+ {1\over 2}(1+Q^2+3y^2)\nonumber \\ &&
+{1\over x+1}\bigg[ e_1^2(Q-y)+e_1 e_2(Q+3)-{1\over 2}(Q+1)^2\bigg]-4e_2^2+Q+2y
+2 Qy \nonumber \\ && + {y\over x} (20e_1^2-4e_1 e_2-4Q+y) +e_1^2(Q+y)+e_1 e_2
(1-6y-3Q) \nonumber \\ &&+ \bigg[{y\over x}(4e_1 e_2+4Q-y-20e_1^2) +{x\over 2}
(2e_1 e_2+2e_2^2-2-Q-3y)-{y\over 2}(7+3y) \nonumber \\ &&+{1\over  2(x+1)^2}
\Big(2e_1^2 (Q-y) +2e_1 e_2(3+Q)-(Q+1)^2\Big)+{1\over x+1}\bigg(2e_1^2(y-3Q)
\nonumber \\ && +e_1 e_2(5y-Q-17)+1-e_2^2+{3\over 2}(3Q+y)+Q(3Q+y-5e_2^2)\bigg) 
-2Q y\nonumber \\ && -e_1^2(Q+y)+e_1 e_2(2Q+5y-6)+e_2^2(5+Q)+{1\over  2}
(Q-Q^2-1) \bigg] \ln(-x)\nonumber \\ &&  + \bigg[ {1\over x}\Big(2e_1 e_2(Q-2
+4y)-2e_1^2(Q+3y) +4e_2^2-2y-Q y-2y^2\Big)\nonumber \\ &&  +{y\over x^2}(4e_1
e_2+4Q-y- 20e_1^2)+2e_1 e_2-2y-1\bigg] \bigg[{\rm Li}_2(x+1)-{\pi^2\over 6} 
\bigg]  \bigg\}\,.\end{eqnarray}

%\medskip
\noindent{\it Class IV.} The vertex correction to the (lower) virtual photon
absorption introduces loop functions that depend on two variables, $x$ and $Q$. 
The pertinent interference terms with the tree diagrams ($A$ and $B$) read:
\begin{eqnarray} {\rm IV}\otimes A &=& {\alpha \over 4\pi x}\Bigg\{-2A\otimes A
\,\bigg[\xi_{UV}+\int_0^1\!\!da\!\int_0^1\!\!db\,b\ln K(a,b;x,Q)\bigg]+e_1^2(6-5x) 
+5e_1e_2 x\nonumber \\ && -{1\over 4}(6Q+5x y) +\int_0^1\!\!da\!\int_0^1\!\!db\,
{1\over K(a,b;x,Q)}\bigg\{ 2e_1^2(8+4Q-Qx+xy)\nonumber \\ && +2e_1 e_2 x(2+Q+x)-
2Q^2-Q(4+3x+ x y)-x(2x+y+xy)\nonumber \\ && +\Big[2e_1^2(Q x+4x-8-4Q-x y)
-2e_1 e_2 x(4+Q+x)+Q(8+3x+xy) \nonumber \\ && +2Q^2+x(2x+3y+xy)\Big] b + 
\Big[2e_1 e_2(Q-x)-2e_1^2(Q+y)+2Q+xy \Big] x\, ab \nonumber \\ &&+ \Big[2e_1^2
(4Q-Qx+xy) +(2e_1 e_2 x-2Q -xy)(Q+x)\Big]ab^2\nonumber \\ && -\Big[4e_1^2(2+x)
+2Q+xy\Big] b^2 + \Big[4e_1^2(x-2)-4e_1 e_2 x+2 Q + x y\Big] Q\, a^2b^2\bigg\}
\Bigg\}\,, \end{eqnarray}

\begin{eqnarray}{\rm IV}\otimes B &=& {\alpha \over 4\pi x}\Bigg\{-2A\otimes B
\,\bigg[\xi_{UV}+ \int_0^1\!\!da\!\int_0^1\!\!db\,b\ln K(a,b;x,Q)\bigg]+{5e_1^2
\over 2}(Q+y)\nonumber \\ && +{5e_2^2\over 2} (Q+x) +{e_1 e_2\over 2}(12-4Q-5x
-5 y) -{x y\over 2}-{3 Q\over 4}(2+Q+x+y)\nonumber \\ && +\int_0^1\!\!da\!
\int_0^1\!\!db\, {1\over K(a,b;x,Q)}\bigg\{2e_1^2(2+Q)(Q+y)+2e_2^2(2Q+Q^2+2Qx
+x^2) \nonumber \\ && + 2e_1 e_2 (8+4Q+2x-2y -Q y + x y)-Q^3+ x y - Q^2(4+2x+y)
\nonumber \\ &&  - Q (4 + 3x + x^2  + 2y+x y)+\Big[ Q^3-xy
-2e_1^2(2+Q)(Q+y)\nonumber \\ && + 2 e_1 e_2(2x+2y-8-2Q+ Q y - x y)-2e_2^2
(4Q+Q^2+2x+2Qx + x^2) \nonumber \\ && +Q^2(6+2x+y)+Q(8+5x+x^2+4y+ x y)
\Big] \,b + \Big[ Q(2 + Q + x + y) \nonumber \\ && -2 e_2^2 (Q + x) - 2 e_1 e_2 
(Q + y)\Big]\,x\,ab+ \Big[2e_1^2Q(Q+y)+2 e_1 e_2 (4Q-Q y+x y)\nonumber \\ &&
+2 e_2^2(Q+x)^2-Q^3-Q x (2 + x + y) - Q^2(2+2x+y)\Big]\,a b^2 + \Big[2e_2^2(Q+x) 
\nonumber \\ &&-2e_1 e_2(4+2Q+3x-y) -2e_1^2(Q+y)-Q^2-Q(2+x+y) \Big] \,b^2 
+Q\,a^2b^2\nonumber \\ &&  \times  \Big[2e_1 e_2 (x+y-4) -2 e_1^2(Q+y)-2e_2^2
(Q+ x)+Q(Q+2+x+y) \Big] \bigg\} \Bigg\}\,, \end{eqnarray}
with the cubic polynomial $K(a,b;x,Q) = b+x a (b-1)+Q a(1-a)b$ in two Feynman
parameters $a, b$. In the physical region, $x>0$ and $Q>0$, the integrals 
involving $1/K(a,b;x,Q)$ are singular and in fact they become complex-valued. 
In order to compute their real parts (which are only of relevance according to 
eq.(31)), one performs the Feynman parameter integral $\int_0^1 db$ 
analytically and converts the occurring (complex) logarithm $\ln(-a x)$ into 
$\ln(a x)$. The remaining integral $\int_0^1 da$ is free of poles and 
therefore unproblematic for a numerical treatment. For the contributions of 
the crossed photon-loops, where $x>0$ is replaced by $y<0$, the two-parameter 
integrals in eqs.(38,39) are real-valued and can be evaluated numerically as 
they stand.   

On the mass-shell the vertex corrections arising from the diagrams III and IV 
reduce to the anomalous magnetic moment $\kappa= \alpha/2\pi$. This piece
alone leads to the following interference terms with the tree diagrams:
\begin{equation} \kappa \otimes A = {\alpha \over 4\pi } \bigg\{ 3e_1^2-3e_1
  e_2+{Q \over 2x}(x+4)+{1\over 4}(7x+5y)\bigg\} \,, \end{equation}
\begin{eqnarray}  \kappa \otimes B &=& {\alpha \over 4\pi x } \bigg\{e_1
  e_2(4Q+x+2y)-e_1^2(Q+2y)-e_2^2(3Q+x) \nonumber \\ && \qquad \,\, +Q\bigg(2+
{x\over 2}+y \bigg) +Q^2-{y\over 4}(3x+y)\bigg\}\,. \end{eqnarray} 

%\medskip
\noindent{\it Class V.} The box diagram is most tedious to evaluate. A good 
fraction of the occurring loop integrals with three and four propagators
can be still be solved in closed analytical form, while the remaining ones have
to be represented as parametric integrals. Putting all pieces together, we
find the following rather lengthy expressions for the interference terms of
the photon-loop diagram V with the tree diagrams ($A$ and $B$):  
\begin{eqnarray}{\rm V}\otimes A &=& {\alpha \over 4\pi}\Bigg\{\Big[4e_1^2
(Q-4+3x+2y)+4e_1e_2(4-3Q-4x-3y)+(x+y)(2+x+y)\nonumber \\ &&+Q(2Q+6+3x+2y)\Big]
{1\over x+y} \Big[ (Q+x+y)\, L^2(Q+x+y)-Q\,L^2(Q) \Big] \nonumber \\ &&+
\Big[8e_1^2(2-2x-y)+4e_1e_2(4x-4+3y) +4e_2^2 x-2Qx-(2+x)(x+y)\Big] \nonumber \\
&& \times {1\over (x+y)^2} \bigg[ x+y+ (Q+x+y)^2\, L^2(Q+x+y)-Q(Q+x+y)\,
L^2(Q)\nonumber \\ && + (2Q+x+y) \sqrt{4+Q}\, L(Q)-2(Q+x+y) \sqrt{4+Q+x+y}\, 
L(Q+x+y)\bigg] \nonumber \\ &&+\Big[2e_1^2(8-3Q-6x-5 y)+2 e_1 e_2 (8 Q-8+7x+7y)
+2e_2^2(x-Q)-2(x+y) \nonumber \\ && - Q(2Q+4+3x+y)\Big] {1\over x+y}\bigg[
\sqrt{4+Q+x+y}\,  L(Q+x+y)-\sqrt{4+Q}\, L(Q)\bigg]\nonumber \\ &&+(2e_1^2+2e_1
e_2-Q-x) {1\over x+y} \bigg[ 2 (Q+x+y)\, L^2(Q+x+y)-2Q\,L^2(Q)\nonumber \\ &&
-{3\over 2}(x+y)+(Q+x+y) \sqrt{4+Q+x+y}\,  L(Q+x+y)-Q \sqrt{4+Q}\, L(Q)\bigg]
\nonumber \\ && +\Big[ e_1(x+y)+2e_2(Q+x)\Big] {e_1-e_2 \over (x+y)^2} \bigg[ 
4(Q+x+y)\, L^2(Q+x+y)-4Q\,L^2(Q)\nonumber \\ && -x-y +2(Q+x+y)\bigg( \sqrt{
4+Q+x+y}\,  L(Q+x+y)-\sqrt{4+Q}\, L(Q)\bigg)\bigg]   \nonumber \\ && +
\Big[e_1(x-Q)+e_2(Q+x)\Big]  {2(e_1-e_2) \over (x+y)^3} \bigg[2(Q+x+y)(Q+x+y-2)
\nonumber \\ && \times \Big[ (Q+x+y)\,L^2(Q+x+y)-Q\,L^2(Q)\Big]+(x+y)(3Q+4x+4y) 
\nonumber \\ &&-6(Q+x+y)^2 \sqrt{ 4+Q+x+y}\, L(Q+x+y)+\Big(6Q^2+10Q(x+y)
\nonumber \\ && + 3(x+y)^2 \Big) \sqrt{4+Q}\, L(Q)\bigg]+ {e_1^2 \over x+1}
+{4\over x} (4e_1e_2-e_1^2-3e_2^2) +6e_1^2\nonumber \\ && -8e_1e_2 -4e_2^2
+2Q+2y+2+x+ \bigg[{4\over  x}(e_1^2-4e_1e_2+3e_2^2) +{e_1^2  \over 
(x+1)^2} \nonumber \\ &&+{Q+1-4e_1^2+4e_1e_2 \over x+1} -3e_1^2+2e_1 e_2+4e_2^2-1
-Q-y \bigg] \ln(-x) \nonumber \\ && +\Big[2e_1^2(2-x^2)+2e_1 e_2(x^2-3x-8)+2e_2^2
(6+5x) -2Qx-2x y-{x^2 y\over 2} \Big]\nonumber \\ && \times {1\over x^2}
\bigg[{\rm Li}_2(x+1) -{\pi^2\over 6} \bigg] 
+(Q+x-2e_1^2) \bigg[-{1\over  4}+\int_0^1\!\!da\!\int_0^1
\!\!db\,b\ln K(a,b;x,Q)\bigg] \nonumber \\ &&+\int_0^1\!\!da\!\int_0^1\!\!db\, 
{1\over K(a,b;x,Q)}\bigg\{2e_1^2(2-2x-y)+2e_1e_2 x+{Q\over 2}(y-x-2)\nonumber 
\\ && -{x\over 2}(x+y)+  \Big[2e_1^2(2x+y)-2e_1e_2(Q+x)+{Q^2\over 2}+Q(x+1)
+{x\over 2}(2+x) \Big] \,b \nonumber \\ && +{x-Q \over 2}(4e_1^2+Q+x) \, a b+ 
2e_1^2(Q-x) \, ab^2-4e_1^2\,b^2 \bigg\}- A\otimes A \,{2+Q+x+y\over x(\tilde
h_+ -\tilde h_-)} \nonumber \\ && \times \Big[8\xi_{IR}\sqrt{Q+x+y}\, L(Q+x+y)
+2{\rm Li}_2(\tilde w)- 2{\rm Li}_2(1-\tilde w) +\ln^2\tilde w-\ln^2(1-\tilde w)
\Big]  \nonumber \\ && +\Big[4e_1^2 (1+Q+x+y)+4e_1e_2(Q+x+y-2)-12e_2^2-Q(Q+6+
3x+y) \nonumber \\ && -(4+2x)(2+x+y)\Big]\,{L(Q+x+y) \over \sqrt{4+Q+x+y}}+
\Big[2e_1^2(x-4)-2e_1e_2 x+2Q +{x y\over 2}\Big]\nonumber \\ &&\times (2+Q+x+y)
{1\over x}\bigg[{{\rm Li}_2(\tilde h_+)- {\rm Li}_2(\tilde h_-)\over\tilde h_+ 
-\tilde h_-}+ \int_0^1\!\!da {\ln[1+a(1-a)Q]\over 1+a(1-a)(Q+x+y)}\nonumber \\ 
&&-{4\ln(-x) \,L(Q+x+y)\over \sqrt{4+Q+x+y}} \bigg]+\int_0^1\!\!da\!\int_0^1\!\!
db\!\int_0^1\!\!dc {2a b\over G^2(a,b,c;x,y,Q)} \bigg\{\Big[e_2x(Q+x\nonumber 
\\ && +y-1) +e_1(8+4Q+3x+4y)\Big]\Big[e_1 (1-ac)+e_2\,ac\Big] -\Big[x(1+Q+x+y) 
\nonumber \\ && +2Q+3y\Big] \bigg[b\Big(e_1(1-ac)+e_2\,ac\Big)^2- {1\over 2} 
G(a,b,c;x,y,Q)\bigg] \bigg\} \Bigg\}\,, \end{eqnarray}

\begin{eqnarray}{\rm V}\otimes B &=& {\alpha \over 4\pi}\Bigg\{\Big[8e_2^2
(1-Q-x-y)+2e_1e_2(3Q+3x+4y-4)-2e_1^2(Q+y)+x+y \nonumber \\ &&+Q(2Q+3+2x+2y)\Big]
{2\over x+y} \Big[ (Q+x+y)\, L^2(Q+x+y)-Q\,L^2(Q) \Big] \nonumber \\ &&+
\Big[4e_1^2(Q+y)+2e_1e_2(4-5Q-4x-5y) +2e_2^2(3Q-4+5x+4y) -Qx\nonumber \\ && -x-y
\Big] {2\over (x+y)^2} \bigg[ x+y+ (Q+x+y)^2\, L^2(Q+x+y)-Q(Q+x+y)\,
L^2(Q)\nonumber \\ && + (2Q+x+y) \sqrt{4+Q}\, L(Q)-2(Q+x+y) \sqrt{4+Q+x+y}\, 
L(Q+x+y)\bigg] \nonumber \\ && +\Big[3e_1^2(Q+y)+e_1 e_2 (8-8 Q-7x-11y)
+e_2^2(9Q+9x+10y-8)- x-y \nonumber \\ && -Q (2+Q+x+y)\Big] {2\over x+y}\bigg[
\sqrt{4+Q+x+y}\,  L(Q+x+y)-\sqrt{4+Q}\, L(Q)\bigg]\nonumber \\ &&+(4e_2^2-Q-y) 
{1\over x+y} \bigg[ 2 (Q+x+y)\, L^2(Q+x+y)-2Q\,L^2(Q)\nonumber \\ &&
-{3\over 2}(x+y)+(Q+x+y) \sqrt{4+Q+x+y}\,  L(Q+x+y)-Q \sqrt{4+Q}\, L(Q)\bigg]
\nonumber \\ && +\Big[e_2(Q+x+y)-e_1(Q+y)\Big] {2(e_1-e_2) \over (x+y)^2} \bigg[ 
4(Q+x+y)\, L^2(Q+x+y) -x-y\nonumber \\ &&  -4Q\,L^2(Q)+2(Q+x+y)\bigg( \sqrt{
4+Q+x+y}\,  L(Q+x+y)-\sqrt{4+Q}\, L(Q)\bigg)\bigg]   \nonumber \\ && +
\Big[e_2(Q+x)-e_1(Q+y)\Big]  {2(e_1-e_2) \over (x+y)^3} \bigg[2(Q+x+y)(Q+x+y-2)
\nonumber \\ && \times \Big[ (Q+x+y)\,L^2(Q+x+y)-Q\,L^2(Q)\Big]+(x+y)(3Q+4x+4y) 
\nonumber \\ &&-6(Q+x+y)^2 \sqrt{4+Q+x+y}\,L(Q+x+y)+\Big(10Q(x+y)+3(x+y)^2
\nonumber \\ && +6Q^2\Big)\sqrt{4+Q}\, L(Q)\bigg] +2+{Q-3y \over 2}-{2y \over
  x}(Q+y) -e_1^2\bigg( {12\over x^2} (Q+y)+{Q+y\over 1+x}\nonumber \\ &&
+{1\over x}(12+5Q+5y) \bigg) + e_1 e_2\bigg(10+{3+Q \over x+1}+{12\over x^2}
(2Q+y) +{1\over x}(4+12Q+15y)\bigg)  \nonumber \\ &&+{e_2^2 \over  x^2}(8x-12Q
-7Qx-9x^2-4 x y) +\bigg[e_1^2\bigg( {12\over x^2}(Q + y)+{Q + y-6 \over  x+1}
-{Q + y\over (x+1)^2}\nonumber \\ && +{2\over x}(6+Q+y) \bigg)+ e_2^2\bigg(5+ 
{12 Q\over x^2}-{Q+1\over x+1}+{4\over x} (Q+y-2)\bigg) + e_1e_2 \bigg({3+Q\over 
(x+1)^2}\nonumber \\ &&-{12\over x^2}(2Q+y)+{3y-6-2Q\over x+1}-{2\over x}
(2+3Q+6y)-5 \bigg) +{Q+3y\over 2}+{2y \over x}(Q+y) \nonumber \\ && +{5Q+2Q^2
+3y +2Qy\over 2(x+1)} \bigg]\, \ln(-x) +\Big[2e_1^2(6Q+6x+4Qx+3x^2 +Qx^2+6y 
\nonumber \\ && +4xy+x^2y) -2e_1e_2(12Q+2x+9Qx+5x^2+2Qx^2+2x^3+6y+9xy+3x^2y)
\nonumber \\ &&+2e_2^2(6Q-4x +5Qx+x^2+2Qx^2+2x^3+2xy+2x^2y)-Q^2x^2+xy(x+2y) 
\nonumber \\ && +Qx(2y-x-x^2-x y) \Big] {1\over  x^3}\bigg[ {\rm Li}_2(x+1)
-{\pi^2\over 6}\bigg]- A\otimes B \,{2+Q+x+y\over  x(\tilde h_+ -\tilde h_-)}
\Big[ 8 \xi_{IR} \nonumber \\ && \times\sqrt{Q+x+y}\, L(Q+x+y) + 2{\rm Li}_2
(\tilde w)- 2{\rm Li}_2( 1-\tilde w) +\ln^2\tilde w -\ln^2(1-\tilde w)\Big] 
\nonumber \\ && +\int_0^1\!\!da\!\int_0^1\!\!db\, {1\over  K(a,b;x,Q)}\Big\{
4e_1e_2(Q+x+y-1)(b-1)-Q(1+Q+x+y) \Big\}\nonumber \\ && + \Big[6 e_1^2+2 e_1 
e_2(2-Q-x-y)-2 e_2^2(1+Q +x+y)- Q-4-2x \Big] \nonumber \\ && \times {2L(Q+x+y) 
\over \sqrt{4+Q+x+y}}+ \Big[2e_1 e_2(2Qx+2x^2-8-4Q- 2 x- 2 y +Qy+3xy+y^2) 
\nonumber \\ &&  -2e_1^2 (2Q+Q^2+3x+Qx+ 2y + 2 Q y + x y + y^2 )+Q (2 +Q+x+ 
y)^2 \nonumber \\ &&-x^2-2e_2^2(2Q+Q^2+3x+ Q x + Q y)\Big]{1\over x} \bigg[ 
\int_0^1\!\!da{\ln[1+ a(1-a)Q]\over 1+a(1-a)(Q+x+y)}\nonumber \\ &&+{{\rm Li}_2
(\tilde h_+)- {\rm Li}_2(\tilde h_-)\over \tilde h_+- \tilde h_-}-{4\ln(-x) 
\,L(Q+x+y)\over \sqrt{4+Q+x+y}}\bigg] +
\int_0^1\!\!da\!\int_0^1\!\!db\!\int_0^1\!\!dc {2a b\over G^2(a,b,c;x,y,Q)}
\nonumber \\ && \times \bigg\{\Big[e_1(6 Q + 2 Q^2  + 6 x + 2 Q x + 3 y + 4Qy
 +2xy+2y^2)+ e_2 (8+2 Q -2 Q x \nonumber \\ &&+2 x-2 x^2-y-Q y-3 x y-y^2 )\Big]
\Big[e_1(1-ac)+e_2\,ac\Big]-\Big[Q(Q+2+x+2y) \nonumber \\ &&+3x+y+xy+y^2
\Big]\bigg[b\Big(e_1(1-ac)+e_2\,ac\Big)^2- {1\over 2} G(a,b,c;x,y,Q)\bigg]
\bigg\} \Bigg\}\,.\end{eqnarray}
Here, we have introduced the auxiliary variables: 
\begin{equation} \tilde w ={1\over 2} \Bigg( 1-\sqrt{{Q+x+y\over 4+Q+x+y}}\,
\Bigg)\,, 
\qquad \tilde h_\pm ={1\over 2}\bigg[-Q-x-y\pm\sqrt{(Q+x+y)(4+Q+x+y)}\,\bigg]\,,
\end{equation}
and the quartic polynomial $G(a,b,c;x,y,Q)= b +a(b+c-1-abc)\,x+ a(1-a)b
(c\,y+Q)$ in three Feynman parameters $a,b,c$. The logarithmic functions 
$L(Q)$ and $L(Q+x+y)$ are defined by eq.(16). The (solvable) integrals 
$\int_0^1 dc$ over the third Feynman parameter $c$ occur in eqs.(42,43) in the 
following four versions: 
\begin{equation} \int_0^1\!dc \, {-a \over G(a,b,c;x,y,Q)} = {1\over
 N(a,b;x,y)}\ln{K(a,b;x,Q) \over b[1+a(1-a)(Q+x+y)]} \,, \end{equation}
\begin{equation} \int_0^1\!dc \, {b \over G^2(a,b,c;x,y,Q)} = {1\over [1+a(1-a)
(Q+x+y)]K(a,b;x,Q)} \,, \end{equation}
\begin{eqnarray} \int_0^1\!dc \, {-a^2bc \over G^2(a,b,c;x,y,Q)} &=& {a\over
N(a,b;x,y)[1+a(1-a)(Q+x+y)]}\nonumber \\ && +{b\over N^2(a,b;x,y)} 
\ln{K(a,b;x,Q)\over  b[1+a(1-a)(Q+x+y)]}\,, \end{eqnarray}
\begin{eqnarray} \int_0^1\!dc \, {a^3bc^2 \over G^2(a,b,c;x,y,Q)} &=& {2b 
K(a,b;x,Q) \over N^3(a,b;x,y)} \ln{K(a,b;x,Q)\over b[1+a(1-a)(Q+x+y)]}
\nonumber \\ && +{2ab \over N^2(a,b;x,y)}-  {a^2 \over N(a,b;x,y)[1+a(1-a)
(Q+x+y)]} \,, \end{eqnarray}
introducing a new denominator polynomial $N(a,b;x,y) = x(1-a b)+y(1-a)b$. It
is strictly positive for $x>0$ and $x+y>0$. The latter inequality imposes on
the squared invariant momentum transfer $t= (p_1-p_2)^2$, the condition $-t> Q 
m^2 =\vec q\,^2$. As has been discussed in section 3 of ref.\cite{bremsrad} 
this represents no restriction for the experimentally selected events in the 
Coulomb peak, where $Q<0.1$ (or even less). The term with $1/K(a,b;x,Q)$ in 
eq.(46) requires a further analytical integration $\int_0^0 db$ (and 
conversion of logarithms into those of absolute values) before it can be 
handled numerically. 

As a good check one verifies that the ultraviolet divergencies proportional to
$\xi_{UV}$ cancel in the sums (I+II+III+IV)$\,\otimes\, A$ and 
(I+II+III+IV)$\,\otimes\, B$.     
\subsection{Infrared finiteness}
The infrared divergent terms proportional to $\xi_{IR}$ in eqs.(32,33,42,43)
follow again the pattern of the Born terms, $A\otimes A/x$ and $A\otimes B/x$.
The differential cross section $d^5\sigma/d\omega d\Omega_{\gamma} d\Omega_{\mu}$ 
at leading order gets therefore multiplied by the (infrared divergent) 
factor:   
\begin{equation} \delta_{\rm virt}^{(\rm IR)}= {2 \alpha \over \pi} \Bigg[ 1-
{4+2Q+2x+2y\over \sqrt{4+Q+x+y}}\, L(Q+x+y) \Bigg] \, \xi_{IR} \,. \end{equation}
On the other hand, the soft photon radiation from the muon (before and after 
the virtual Compton scattering) yields the multiplicative factor: 
\begin{equation} \delta_{\rm soft}= \alpha\, \mu^{4-d}\!\!\int\limits_{|\vec 
l\,|<\lambda} \!\!{d^{d-1}l  \over (2\pi)^{d-2}\, l_0} \bigg\{ {2p_1\cdot p_2 
\over p_1 \cdot l \, p_2 \cdot l} - {m^2 \over (p_1 \cdot l)^2} - {m^2 \over 
(p_2 \cdot l)^2} \bigg\} \,, \end{equation}
which includes exactly the same infrared divergence $(\sim \xi_{IR})$ but with 
the opposite sign. The remaining finite radiative correction factor reads:  
\begin{eqnarray}\delta^{\rm (lab)}_{\rm real}  &=&{\alpha\over\pi}\Bigg\{\bigg[2-{4
(2+Q+x+y) \over \sqrt{4+Q+x+y}} L(Q+x+y)\bigg] \ln{m \over 2\lambda}
\nonumber \\ && +{e_1 \over \sqrt{e_1^2-1}} \ln\Big(e_1+\sqrt{e_1^2-1}\Big)
 +{e_2 \over \sqrt{e_2^2-1}} \ln\Big(e_2+\sqrt{e_2^2-1}\Big) \nonumber \\ && 
-\int_0^1\!\!d\tau \,{(2+Q+x+y) [\tau e_1+(1-\tau) e_2] \over 2[1+\tau(1-\tau)
(Q+x+y)]\sqrt{R}}\,\ln{\tau e_1+(1-\tau)e_2+\sqrt{R}\over \tau e_1+(1-\tau) 
e_2 -\sqrt{R}} \Bigg\}\,,\end{eqnarray} 
with the abbreviation $R=[\tau e_1+(1-\tau)e_2]^2-1+\tau(\tau-1)(Q+x+y)$.
We note that the terms beyond those proportional to $\ln(m/2\lambda)$ are
specific for the evaluation of the soft photon correction factor 
$\delta_{\rm  soft}$  in the laboratory frame with $\lambda$ an infrared
cut-off therein.  Alternatively, it can also be evaluated in the muon-photon
center-of-mass frame, see herefore the expression $\delta_{\rm real}^{(\rm  cm)}$ 
written in eq.(24) of ref.\cite{bremsrad}. 

Finally, it is important to compare our results with Fomin's earlier 
calculation of the radiative corrections to bremsstrahlung in 
refs.\cite{fomin1,fomin2}. For the soft photon contribution 
$\delta^{\rm  (lab)}_{\rm real}$, written here in eq.(51), the agreement is obvious by 
direct comparison with the expressions given in eqs.(42-45) of 
ref.\cite{fomin2}. Since Fomin uses a fictitious photon mass $m_\gamma$ to 
regularize infrared divergencies our quantity $\xi_{IR}$ is to be identified 
with the logarithm $\ln(m/m_\gamma)$. The radiative corrections due to photon
loops are summarized in Fomin's work by the term $\delta_{\rm virt}= -{\alpha \over
 \pi}[U^{re}(-x,-y,Q,e_1,e_2)+U^{re}(-y,-x,Q,e_2,e_1)]/U_0$ with $U_0=H_{\rm tree}$ 
(see eqs.(26-29)) and the function $2U^{re}(-x,-y,Q,e_1,e_2)$ is specified in
eqs.(36-41) of ref.\cite{fomin2}.  We have evaluated the term $\delta_{\rm virt}$ 
(setting $\xi_{IR}=0$) numerically at various points in the physical region 
$x,Q>0$, $y<0$, $e_{1,2}>1$  and obtained agreement with our result $H_{\rm
loop}/H_{\rm  tree}$ (see eq.(31)) with good numerical precision. Another 
radiative correction (called $-2\alpha W/\pi$ by Fomin \cite{fomin2}) which 
routinely is included in the bremsstrahlung process is the (leptonic) vacuum 
polarization to the virtual photon exchange. In our notation the radiative 
correction factor due to muonic vacuum polarization reads:
\begin{equation} \delta_{\rm vp,\mu} = {4 \alpha \over 3 \pi} \bigg\{ {2\over    Q} 
-{5 \over 6}+\bigg( 1-{2\over Q}\bigg)\sqrt{4+Q}\, L(Q)\bigg\} \,,\end{equation}
with $L(Q)$ defined by eq.(16). More significant is of course the additional 
radiative correction factor $\delta_{\rm vp,e}$ from electronic vacuum 
polarization which is obtained simply through the substitution $ Q\to 
Q(m/m_e)^2= 42753 Q$ in eq.(52).

The analytical results presented in section 3 of this paper form the basis for 
a detailed study of the radiative corrections to muon-nucleus bremsstrahlung 
$\mu^- Z \to \mu^- Z \gamma$. As mentioned in the introduction  muon-nucleus
bremsstrahlung  serves in the COMPASS experiment \cite{compass} at CERN as a 
calibration process for pion-nucleus bremsstrahlung $\pi^- Z \to \pi^- Z 
\gamma$ with the aim of measuring with high statistics the pion electric and 
magnetic  polarizabilities. 

Before closing the paper let us add a few remarks. Naturally, one expects that 
the radiative corrections are approximately equal for electromagnetic 
processes with muons and pions. This expectation is to some degree confirmed by 
our comparison of the radiative corrections to muon and pion Compton 
scattering (of real photons) in section 2.3 (see Figs.\,4,5). The dominance of 
the real radiative corrections (due to soft photon radiation) over the virtual 
radiative corrections (due to photon-loops) for small enough infrared
cut-offs $\lambda$ gives an argument in the same direction. The contribution
scaling with the logarithm $\ln(m/2\lambda)$ is in fact universal.    
   
\section*{Acknowledgement}
I thank Jan Friedrich for triggering this work and for many informative
discussions.

\end{document}